\documentstyle[preprint,aps,tighten]{revtex}
\begin{document}
\draft

%\textheight=23 truecm
%\textwidth=16.0 truecm
%\topmargin=-1.0cm
%\evensidemargin=0.0cm
%\oddsidemargin=0.0cm
\tolerance=5000
\def\pp{{\, \mid \hskip -1.5mm =}}

\def\cL{{\cal L}}
\def\be{\begin{equation}}
\def\ee{\end{equation}}
\def\bea{\begin{eqnarray}}
\def\eea{\end{eqnarray}}
\def\tr{{\rm tr}\, }
\def\nn{\nonumber \\}
\def\e{{\rm e}}
\def\D{{D \hskip -3mm /\,}}

\def\SEH{S_{\rm EH}}
\def\SGH{S_{\rm GH}}
\def\AdS5{{{\rm AdS}_5}}
\def\S4{{{\rm S}_4}}
\def\gfv{{g_{(5)}}}
\def\gfr{{g_{(4)}}}
\def\SC{{S_{\rm C}}}
\def\RH{{R_{\rm H}}}

%%%%%%%%%%%%%%%%%%%%%%%%%%%%%%%%%%%%%%%%%%%%%%%%%%%%%%%%%55

\title{
STRING VERSUS EINSTEIN FRAME IN  AdS/CFT INDUCED 
QUANTUM DILATONIC BRANE-WORLD UNIVERSE}

\author{Shin'ichi Nojiri\thanks{Electronic address: 
nojiri@cc.nda.ac.jp}}
\address{Department of Applied Physics,
National Defence Academy,
Hashirimizu Yokosuka 239-8686, JAPAN}

\author{Octavio Obregon\thanks{Electronic Address: 
octavio@ifug3.ugto.mx}, 
Sergei D. Odintsov\thanks{On leave from Tomsk State 
Pedagogical University, 
634041 Tomsk, RUSSIA.
Electronic address: odintsov@ifug5.ugto.mx}
and Vladimir I. Tkach\thanks{Electronic Address: 
vladimir@ifug3.ugto.mx}}
\address{
Instituto de Fisica de la Universidad de Guanajuato,
Lomas del Bosque 103, Apdo.\ Postal E-143, 
37150 Leon, Gto., MEXICO }

\maketitle
\begin{abstract}  

AdS/CFT induced quantum dilatonic brane-world where 4d boundary is 
flat or de Sitter (inflationary) or Anti-de Sitter brane is 
considered. The classical brane tension is fixed but boundary 
QFT produces the effective brane tension via the account of 
corresponding conformal anomaly induced effective action. 
This results in inducing of brane-worlds in accordance with 
AdS/CFT set-up as warped compactification. The explicit, 
independent construction of quantum induced dilatonic 
brane-worlds in two frames: string and Einstein one is done. 
It is demonstrated their complete equivalency for all quantum 
cosmological brane-worlds under discussion, including several
examples of classical brane-world black holes. This is different 
from quantum corrected 4d dilatonic gravity where de Sitter 
solution exists in Einstein but not in Jordan (string) frame.
The role of quantum corrections on massive graviton perturbations 
around Anti-de Sitter brane is briefly discussed.

\end{abstract}


\section{Introduction}

Brane-worlds are alternative to the standard Kaluza-Klein 
compactification. They naturally lead to the following nice 
features of mutli-dimensional theory like trapping of 4d 
gravity on the brane \cite{RS}, resolution of hierarchy 
problem and possibly resolution of cosmological constant problem. 
Different aspects of brane-world cosmology (for very incomplete 
list of references see \cite{CH,cosm}) are under very active 
investigation. 

The essential element  of original brane-world models is the 
presence in the theory of two free parameters (bulk cosmological 
constant and brane tension, or brane cosmological constant). 
These parameters are fine-tuned (up to some extent) in order to
construct the successful classical brane-world. This is most 
standard prescription which may be not completely satisfactory 
if one wishes to have the dynamical mechanism of brane tension 
origin.

 From another side, one can fix the classical action on AdS-like 
space from the very beginning with the help of surface terms 
added in accordance with AdS/CFT correspondence\cite{AdS}. 
Such terms should make the variational procedure to be 
well-defined and also they should eliminate the leading divergence 
of the action. Brane tension is not considered as free parameter 
anymore but it is fixed by the condition of finiteness of 
spacetime when brane goes to infinity. In this case, as 
parameters are fixed the consistent brane-world scenario is 
impossible, as a rule. However, other parameters may improve 
the situation when quantum effects are taken into account.
Taking quantum CFT (including quantum gravity!) on the brane
one adds its contribution (the corresponding conformal anomaly
induced effective action) to the total action.
As a result, it changes the brane tension, the quantum induced 
brane-world occurs as it has been discovered in 
refs.\cite{NOZ,HHR}. Actually, this represents the embedding 
of warped compactification (brane-worlds) to AdS/CFT correspondence,
hence one gets AdS/CFT induced quantum brane-worlds \cite{NOZ,HHR} 
where 4d boundary may be flat or de Sitter or Anti-de Sitter 
spacetime. This is clearly the dynamical mechanism to get curved 
brane-world. It is easily generalized for the presence of 
non-trivial dilaton, i.e. AdS/CFT induced quantum dilatonic 
brane-worlds occur\cite{NOO}. In other words, brane-worlds 
are the consequence of the presence of quantum fields on the 
brane in accord with AdS/CFT set-up. Moreover, such induced 
dilatonic brane-worlds are even more related with AdS/CFT 
correspondence as 5d dilatonic gravity represents the bosonic
sector of 5d gauged supergravity (special parametrization).
Even more, the dynamical determination of 4d dilaton occurs.

In the study of quantum induced brane-worlds, in the same way as 
for any other dilatonic gravity the following question appears:
which frame to work with is the physical one?
There are two convenient frames: string (or Jordan) one where scalar 
curvature explicitly couples with dilaton and Einstein frame 
where scalar curvature does not couple with dilaton. Basically 
speaking, one should expect that results obtained in these two 
frames are not equivalent. 

Indeed, in QFT the choice of different variables and (or) form of 
action corresponds to different parametrizations. QFT results 
are parametrization dependent, only S-matrix is gauge and 
parametrization independent. (Even the quantization procedure 
(for review, see \cite{GL}) is parametrization dependent.) 
As usually the consideration is one-loop ,one should expect in 
many cases the explicit parametrization dependence. Moreover, 
it is known that even for classical dilatonic gravity the 
(singular) solution may exist in only one parametrization. 
Hence, the question of frame dependence should be carefully 
analyzed for all solutions at hands. This is the main purpose 
of the present work: to compare string frame quantum induced 
dilatonic brane-worlds with their analogs in Einstein frame.

In the next section as the simple example, 4d dilatonic 
(Brans-Dicke) theory with large $N$ quantum spinor corrections 
is considered. In the Einstein frame where spinor is dilaton 
coupled one the de Sitter Universe solution with decaying 
dilaton exists. Working with the same theory in string (Jordan)
frame where spinor is getting minimal, one finds that above 
solution does not exist. Hence, it is shown that two frames in 
4d dilatonic gravity with quantum corrections are not equivalent.

In third section we consider 5d dilatonic gravity action with 
4d boundary term induced by conformal anomaly of brane, dilaton 
coupled spinor. Explicit examples of de Sitter, flat and 
Anti-de Sitter dilatonic branes are constructed in Einstein 
frame. The dynamical mechanism to determine the dilaton on 
the brane is presented. In section four the same investigation 
is done in string frame. Brane spinor is now minimal. The same 
AdS/CFT induced quantum brane-worlds are proven to exist. Hence, 
for quantum corrected cosmological dilatonic brane-worlds one 
has the equivalency of string and Einstein frames.

In fifth section the equivalency of string and Einstein frames is
demonstrated for number of classical dilatonic brane-world black 
holes.
In section six some remarks on massive graviton modes around
dilatonic AdS4 brane are made. The role of brane quantum 
corrections for massive graviton modes is clarified. 
Brief summary and some outlook are given in final section.

\section{Jordan and Einstein frames for 4d quantum corrected 
dilatonic gravity}

In the study of dilatonic gravities the interesting question appears:
which frame among few possible ones is the physical one?
Basically speaking, there are two convenient frames to work with:
string (or Jordan) frame and Einstein frame. These two are related
by conformal transformation. The best known example is provided by 
the standard Brans-Dicke theory (with matter). The 
4-dimensional action in the Jordan frame is:
\be
S_{BD}=\frac{1}{16\pi}\int d^4x \sqrt{-g} \left[ 
\phi R -\frac{\omega}{\phi}
(\nabla_\mu \phi)(\nabla^\mu \phi) \right]+S_M\,,
\label{bdj}
\ee
where $\phi$ is the Brans-Dicke (dilaton) field with
$\omega$ being the coupling constant and $S_M$ is the matter action. 

Performing the following conformal transformation
and a redefinition of the scalar field
\be
\label{phis}
\tilde{g}_{\mu \nu}=G\phi g_{\mu \nu}\ , \quad
\tilde{\phi}=\sqrt{{2\omega+3} \over 16\pi G} \ln 
\left(G\phi\right)\,,\,\,\,\,
2\omega+3 >0\,. 
\ee
one gets the action in the Einstein frame 
\be
S=\int d^4x \sqrt{-\tilde{g}(x)} \left[\frac{\tilde{R}}{16\pi G}
-\frac{1}{2}
(\tilde{\nabla}_\mu \tilde{\phi})(\tilde{\nabla}^\mu \tilde{\phi})+
\exp{(A\tilde{\phi})}L_M(\tilde{g}) \right]
\,,
\label{bde}
\ee
where $A=-8\sqrt{\frac{\pi G}{2\omega+3}}$. It is expected 
that these two actions (at least for regular solutions) should 
lead to equivalent results.
However, the explicit consideration shows that it is not always so
(for a review, see \cite{FGT}). That is why it was argued in 
ref.\cite{FGT} that it is Einstein frame which is physical one. 
Of course, such state of affairs is not satisfactory.

In quantum field theory the  choice of different variables 
corresponds to different parametrizations. It is known that 
generally speaking it leads to parametrization dependent results: 
it is only S-matrix should be the same in different 
parametrizations. Of course, this should be true only in complete
theory where account of all loops is taken. As usually the 
consideration is one-loop, one should expect parametrization 
dependence already at one-loop.
 
Let us consider the explicit example in Einstein frame where 
quantum corrections are taken into account.
As matter Lagrangian we take the one associated with $N$ 
massless (Dirac) spinors, i.e. 
$L_M=\sum_{i=1}^N \bar\psi_i\gamma^\mu\nabla_\mu\psi^i$. 
There is no problem to add other types of matter 
(say scalar or vector fields).
The above choice is made only for the sake of simplicity.

We shall make use of the 
EA formalism (for an introduction, see \cite{BOS}). 
The corresponding 4d anomaly--induced EA for dilaton coupled
scalars, vectors and spinors has been found in Refs. \cite{NO}.

Hence, starting from the theory with the action (no classical
 background spinors)
\be
S=\int d^4x \sqrt{-g} \left[\frac{R}{16\pi G}
-\frac{1}{2}
(\nabla_\mu \phi)(\nabla^\mu \phi)+
\exp{(A\phi)}\sum_{i=1}^N 
\bar\psi_i\gamma^\mu\nabla_\mu\psi^i  \right]\,,
\label{bde1}
\ee
we will discuss FRW type cosmologies
\be
ds^2=-dt^2+a(t)^2 dl^2\,,
\label{st4}
\ee
where $dl^2$ is the line metric element of a 3-dimensional 
flat space.  

The computation of the anomaly--induced EA for the dilaton 
coupled spinor field has been done in \cite{NO}, and
the result, in the non-covariant local form, reads: 
\bea
\label{vii}
\lefteqn{\hspace{-.8cm}
W=\int d^4x \sqrt{-\bar g} \bigg\{b \bar F \sigma_1
+ 2b' \sigma_1\Big[ \bar{\Box}^2
+ 2 \bar R^{\mu\nu}\bar\nabla_\mu\bar\nabla_\nu 
%\right. \nn && \left. 
- {2 \over 3}\bar R\bar{\Box}
+ {1 \over 3}(\bar\nabla^\mu\bar R)\bar\nabla_\mu
\Big]\sigma_1}
\nn
&&\!\!\!\!\!\!\! %\qquad
+\, b'\sigma_1\Big(\bar G -{2 \over 3}\bar\Box\bar R\Big) 
%\nn && 
-{1 \over 18}(b + b')\left[\bar R - 6 \bar{\Box} \sigma_1
- 6(\bar\nabla_\mu \sigma_1)(\bar\nabla^\mu \sigma_1)
\right]^2\bigg\}\,,
\eea
where $\sigma_1=\sigma+ A\phi /3$, 
the square of the Weyl tensor is given by
$ F= R_{\mu\nu\rho\sigma}R^{\mu\nu\rho\sigma}
-2 R_{\mu\nu}R^{\mu\nu} + {1 \over 3}R^2 $
and Gauss-Bonnet invariant is
$G=R_{\mu\nu\rho\sigma}R^{\mu\nu\rho\sigma}
-4 R_{\mu\nu}R^{\mu\nu} + R^2$.
For Dirac spinors
$b={3N \over 60(4\pi)^2}$, $b'=-{11 N \over 360 (4\pi)^2}$.

Then we find the following Einstein frame, quantum-corrected solution
 whose metric
 is expressed in Jordan frame as
\bea
\label{16}
ds_J^2&=&a_J^2(\eta)\left(-d\eta^2 + dl^2 \right) \nn
a_J^2(\eta)&=&\e^{-{\phi \over \sqrt{2\omega 
+ 3 \over 16G}}}a^2(\eta) \nn
&=& a_0\eta^{-2\zeta} \nn
\zeta&\equiv& {1 \over 2H_1}\sqrt{16\pi G \over 2\omega +3} + 1 \nn
&=&{ 1 \over \sqrt{2\omega + 3}\left\{
 - {3 \over 16}\sqrt{2\omega + 3}
\pm \sqrt{{9 \over 256}(2\omega + 3) - {1 \over 6}}\right\}} +1 \nn
&=&-{1 \over 8}\mp \sqrt{{81 \over 64} 
 - {6 \over 2\omega + 3}}\ .
\eea
Here $a_0$ is an arbitrary constant. On the other hand, 
one finds the dilaton field $\phi_J$ in the Jordan frame as 
\be
\label{phiJJ}
\phi=\phi_0\eta^{{1 \over H_1} \sqrt{16\pi G \over 2\omega + 3}}
=\phi_0\eta^{2(\zeta -1)}\ ,
\quad \phi_0= {1 \over a_0 G }\ .
\ee

Let us analyze the equations of motion in the Jordan frame (for 
the form of transformation to string frame see section 5). 
The variations over $\phi$ and 
$\sigma$ give the following equations:
\bea
\label{phieqJ}
0&=&6 \left(\sigma'' + {\sigma'}^2\right)\e^{2\sigma}
 -{\omega{\phi'}^2 \over \phi^2}\e^{2\sigma} 
 - 2\omega\left({\phi'\e^{2\sigma} \over \phi}\right)\ ,\\
\label{sigmaeqJ}
0&=& {2 \over 16\pi} \left(6\left(\sigma'' + {\sigma'}^2\right)
+ {\omega {\phi'}^2 \over \phi}\right)\e^{2\sigma} 
+ {6\left(\e^{2\sigma}\right)''
 - 12 \left(\sigma'\e^{2\sigma}\right)' \over 16\pi } \nn
&& + 4b'\sigma'''' - 4\left(b + b'\right)\left\{\left(
\sigma'' - {\sigma'}^2 \right)'' + 2\left(\sigma'
\left(\sigma'' - {\sigma'}^2\right)\right)'\right\}\ .
\eea
Here $'\equiv {d \over d\eta}$. 
We can check that the solution (\ref{16}) and (\ref{phiJJ}) 
does not satisfy (\ref{phieqJ}). If the solution in the Jordan 
frame would be equivalent to that in the Einstein frame even in the 
quantum level, we should have $\sigma_1=\sigma_J\equiv \ln a_J$ but 
we have $\sigma_1=\sigma + {A\phi \over 3}
=\sigma - {4 \over 3}\ln G\phi_J$ and 
$\sigma_J=\sigma  - {1 \over 2}\ln G\phi_J$. This is an 
origin of the inequivalence.  Thus, it is demonstrated that for the
Universe model under consideration the Jordan and Einstein frames in
4d dilatonic gravity with quantum corrections are not equivalent. 
Different parametrizations lead to different results (parametrization
choice dependence). The physical results are expecting to be the same only
for S-matrix
in full theory (non-perturbative regime).

\section{Inflationary dilatonic brane-world Universe 
in Einstein frame}

In this section we present the review 
of quantum induced dilatonic brane-worlds found in ref.\cite{NOO}.
The model is discussed in Einstein frame and using euclidean notations.
This scenario represents the extension to non-constant dilaton case the
earlier scenario of refs.\cite{NOZ,HHR} where quantum brane-worlds 
were realized in frames of AdS/CFT correspondence, by adding quantum 
CFT on the brane to effective action.

We start with Euclidean signature action $S$ which is the sum of 
the Einstein-Hilbert action $\SEH$ including dilaton 
$\phi$ with potential $V(\phi)={12 \over l^2}+\Phi(\phi)$, 
the Gibbons-Hawking surface term $\SGH$,  the surface 
counter term $S_1$\footnote{We use the following curvature 
conventions:
\begin{eqnarray*}
R&=&g^{\mu\nu}R_{\mu\nu} \nn
R_{\mu\nu}&=& R^\lambda_{\ \mu\lambda\nu} \nn
R^\lambda_{\ \mu\rho\nu}&=&
-\Gamma^\lambda_{\mu\rho,\nu}
+ \Gamma^\lambda_{\mu\nu,\rho}
- \Gamma^\eta_{\mu\rho}\Gamma^\lambda_{\nu\eta}
+ \Gamma^\eta_{\mu\nu}\Gamma^\lambda_{\rho\eta} \nn
\Gamma^\eta_{\mu\lambda}&=&{1 \over 2}g^{\eta\nu}\left(
g_{\mu\nu,\lambda} + g_{\lambda\nu,\mu} - g_{\mu\lambda,\nu} 
\right)\ .
\end{eqnarray*}}
\bea
\label{Stotal}
S&=&\SEH + \SGH + 2 S_1 , \\
\label{SEHi}
\SEH&=&{1 \over 16\pi G}\int d^5 x \sqrt{\gfv}\left(R_{(5)} 
 -{1 \over 2}\nabla_\mu\phi\nabla^\mu \phi 
 + {12 \over l^2}+\Phi(\phi)\right), \\
\label{GHi}
\SGH&=&{1 \over 8\pi G}\int d^4 x \sqrt{\gfr}\nabla_\mu n^\mu, \\
\label{S1dil}
S_1&=& -{1 \over 16\pi G}\int d^4 \sqrt{\gfr}\left(
{6 \over l} + {l \over 4}\Phi(\phi)\right)\ .
\eea 
Here the quantities in the  5 dimensional bulk spacetime are 
specified by the suffices $_{(5)}$ and those in the boundary 4 
dimensional spacetime are specified by $_{(4)}$. 
The factor $2$ in front of $S_1$ in (\ref{Stotal}) is coming from 
that we have two bulk regions which 
are connected with each other by the brane. It is clear 
that above representation corresponds to Einstein frame. 
In (\ref{GHi}), $n^\mu$ is 
the unit vector normal to the boundary. 

\subsection{Bulk solutions}

In this subsection, we find some explicit solutions in the 
bulk space. 

We now assume the metric in the following form
\be
\label{DP1}
ds^2=f(y)dy^2 + y\sum_{i,j=0}^3\hat g_{ij}(x^k)dx^i dx^j, 
\ee
and $\phi$ depends only on $y$: $\phi=\phi(y)$. 
Here $\hat g_{ij}$ is the metric of the Einstein manifold, which is
defined by $r_{ij}=k\hat g_{ij}$, where $r_{ij}$ is 
the Ricci tensor constructed with $\hat g_{ij}$ and 
$k$ is a constant. Then we obtain the following equations 
of motion in the bulk:
\bea
\label{DP2}
0&=&{3 \over 2y^2} - {2kf \over y} - {1 \over 4}\left(
{d\phi \over dy}\right)^2 
- \left({6 \over l^2} + {1 \over 2}\Phi(\phi)\right)f, \\ 
\label{DP3}
0&=&{d \over dy}\left({y^2 \over \sqrt{f}}{d\phi \over dy}\right)
+ \Phi'(\phi)y^2 \sqrt{f}\ .
\eea

It is convenient to introduce the new coordinate $z$
\be
\label{c2b}
z=\int dy\sqrt{f(y)}\ .
\ee
By solving $y$ with respect to $z$, we obtain the warp
factor $l^2\e^{2\hat A(z,k)}=y(z)$. Here one assumes 
the metric of 5 dimensional space time as follows:
\be
\label{metric1}
ds^2=dz^2 + \e^{2A(z,\sigma)}\tilde g_{\mu\nu}dx^\mu dx^\nu\ ,
\quad \tilde g_{\mu\nu}dx^\mu dx^\nu\equiv l^2\left(d \sigma^2 
+ d\Omega^2_3\right)\ .
\ee
where $d\Omega^2_3$ corresponds to the metric of 3 dimensional 
unit sphere. Suppose that $A(z,\sigma)$ can be decomposed 
into the sum of $z$-dependent part $\hat A(z)$ and 
$\sigma$-dependent part and therefore $l^2\e^{2\hat A(z)}
\hat g_{\mu\nu}=\e^{2 A(z,\sigma)}\tilde g_{\mu\nu}$. 
Then for the unit sphere ($k=3$)
\be
\label{smetric}
A(z,\sigma)=\hat A(z,k=3) - \ln\cosh\sigma\ ,
\ee
for the flat Euclidean space ($k=0$)
\be
\label{emetric}
A(z,\sigma)=\hat A(z,k=0) + \sigma\ ,
\ee
and for the unit hyperboloid ($k=-3$)
\be
\label{hmetric}
A(z,\sigma)=\hat A(z,k=-3) - \ln\sinh\sigma\ .
\ee

When $\Phi(\phi)=0$, there exists 
the following AdS-like solution of the equations of 
motion \cite{NOtwo}
\bea
\label{curv2}
ds^2&=&f(y)dy^2 + y\sum_{i,j=0}^{d-1}\hat g_{ij}(x^k)dx^i dx^j \nn
f&=&{d(d-1)  \over 4y^2
\lambda^2 \left(1 + { c^2 \over 2\lambda^2 y^d}
+ {kd \over \lambda^2 y}\right)} \nn
\phi&=&c\int dy \sqrt{{d(d-1) \over
4y^{d +2}\lambda^2 \left(1 + { c^2 \over 2\lambda^2 y^d}
+ {kd \over \lambda^2 y}\right)}}\ .
\eea
Here $\lambda^2={12 \over l^2}$.

When $\Phi(\phi)\neq 0$, by using (\ref{DP2}) and (\ref{DP3}), 
one can delete $f$ from the 
equations and we obtain an equation that contains only the 
dilaton field $\phi$:
\bea
\label{DP4}
0&=&\left\{ {5k \over 2} - {k \over 4}y^2
\left({d\phi \over dy}\right)^2 + \left({3 \over 2}y 
 - {y^3 \over 6}\left({d\phi \over dy}\right)^2 \right)
\left({6 \over l^2} + {1 \over 2}\Phi(\phi)\right)\right\}
{d\phi \over dy} \nn
&& + {y^2 \over 2}\left({2k \over y} + {6 \over l^2} 
+ {1 \over 2}\Phi(\phi)\right){d^2\phi \over dy^2}
+ \left({3 \over 4} - {y^2 \over 8}
\left({d\phi \over dy}\right)^2 \right)\Phi'(\phi)\ .
\eea
We now consider a solvable case where
\be
\label{ts1}
{6 \over l^2} + {1 \over 2}\Phi(\phi)= - {2k \over y}\ .
\ee
The explicit form, or $\phi$ dependence, of $\Phi(\phi)$ can 
be determined after solving the equations of motion as the following
\be
\label{ts4}
\phi =\pm \sqrt{6} \ln (m^2 y)\ .
\ee
Here $m^2$ is a constant of the integration. The 
  explicit form 
of $\Phi(\phi)$ is:
\be
\label{ts5}
\Phi(\phi)= - {12 \over l^2}
 - 4km^2\e^{\mp {\phi \over \sqrt{6}}}\ .
\ee
One can also find that Eq.(\ref{DP2}) is trivially satisfied. 
Integrating (\ref{DP3}), we obtain
\be
\label{ts6}
f={1 \over -{2ky \over 9} + {f_0 \over y^2}}\ .
\ee
Here $f_0$ is a constant of the integration and $f_0$ should be 
positive in order that $f$ is positive for large $y$. 
There is a (curvature) singularity at $y=0$. 
One should also note that when $k>0$, the horizon appears  at
\be
\label{ts7}
y^3 = y_0^3\equiv {9f_0 \over 2k}\ 
\ee
and we find  
\be
\label{ts7b}
y\leq y_0\ .
\ee

\subsection{Brane solutions} 

In this subsection, we investigate if there is a solution 
with brane including the quantum correction from $N$ 
massless brane Majorana spinors coupled with the 
dilaton. For simplicity, only the case that the potential is 
constant. 

On the brane, one obtains the following 
equations by the variations over $A$ and $\phi$: 
\bea
\label{eq2b}
0&=&{48 l^4 \over 16\pi G}\left(\partial_z A - {1 \over l}
 - {l \over 24}\Phi(\phi)\right)\e^{4A}\ , \\
\label{eq2pb}
0&=&-{l^4 \over 8\pi G}\e^{4A}\partial_z\phi
 -{l^5 \over 32\pi G}\e^{4A}\Phi'(\phi) \ .
\eea
With (\ref{ts5}) and the solution 
(\ref{ts6}),  these equations look 
\bea
\label{tsc1}
0&=&{1 \over 2R^2}\sqrt{
{f_0 \over R^4} - {2kR^2 \over 9}} - {1 \over 2l}
+ {kl \over 3R^2}, \\
\label{tsc2}
0&=&\sqrt{{f_0 \over R^4} - {2kR^2 \over 9}} + {kl \over kl}\ .
\eea
Here we assume that the brane lies at $y=y_0$ or $z=z_0$. 
The radius $R$ of the brane is defined by $R=\e^{\hat A(z_0)}$.
Eq.(\ref{tsc2}) tells that $k\leq 0$ but by combining (\ref{tsc1}) 
and (\ref{tsc2}), we find $R^2={kl^2 \over 2}$. Then there is no 
consistent classical solution. 

We now consider the case that the matter on the 
brane is  
some QFT like QED or QCD. Of course, 
such a theory is classically  conformally invariant one. As 
an explicit example in order to be able to apply large 
$N$-expansion we suppose that dominant contribution is due 
to $N$ massless Majorana spinors coupled with the 
dilaton, whose action is given by
\be
\label{SP1}
S=\int \sqrt{\gfr} \e^{a\phi}\sum_{i=1}^N\bar\Psi_i \gamma^\mu
D_\mu\Psi_i\ .
\ee
The case of minimal spinor coupling corresponds to the choice $a=0$.
Note that  from Brans-Dicke theory consideration one knows that
for Einstein frame the non-minimal dilaton coupling 
with the matter is the typical case. 
Then the trace anomaly induced action $W$ has the following form \cite{NO}:
\bea
\label{W2}
W&=& b \int d^4x \sqrt{\widetilde g}\widetilde F A_1 \nn
&& + b' \int d^4x\left\{A_1 \left[2{\widetilde\Box}^2 
+\widetilde R_{\mu\nu}\widetilde\nabla_\mu\widetilde\nabla_\nu 
 - {4 \over 3}\widetilde R \widetilde\Box^2 
+ {2 \over 3}(\widetilde\nabla^\mu \widetilde R)\widetilde\nabla_\mu
\right]A_1 \right. \nn
&& \left. 
+ \left(\widetilde G - {2 \over 3}\widetilde\Box \widetilde R
\right)A_1 \right\} \\
&& -{1 \over 12}\left\{b''+ {2 \over 3}(b + b')\right\}
\int d^4x \left[ \widetilde R - 6\widetilde\Box A_1 
 - 6 (\widetilde\nabla_\mu A_1)(\widetilde \nabla^\mu A_1)
\right]^2 \ .\nonumber
\eea
Here 
\be
\label{SP2}
A_1=A+{a\phi \over 3},
\ee
and 
\be
\label{SP3}
b={3N \over 60(4\pi)^2}\ ,\quad b'=-{11 N \over 360 (4\pi)^2}\ .
\ee
We also choose $b''=0$ as it may be changed by finite renormalization 
of classical gravitational action.
%%%%%%%
In (\ref{W2}), one chooses 
the 4 dimensional boundary metric as 
\be
\label{tildeg}
\gfr_{\mu\nu}=\e^{2A}\tilde g_{\mu\nu},
\ee 
and we specify the 
quantities given by $\tilde g_{\mu\nu}$ by using $\tilde{\ }$. 
$G$ ($\tilde G$) and $F$ ($\tilde F$) are the Gauss-Bonnet
invariant and the square of the Weyl tensor, which are given as
\bea
\label{GF}
G&=&R^2 -4 R_{ij}R^{ij}
+ R_{ijkl}R^{ijkl}, \nn
F&=&{1 \over 3}R^2 -2 R_{ij}R^{ij}
+ R_{ijkl}R^{ijkl} \ ,
\eea

For simplicity, we consider a constant potential  
($\Phi(\phi)=0$) case. Then brane equations are 
\bea
\label{eq2c}
0&=&{48 l^4 \over 16\pi G}\left(\partial_z A - {1 \over l}
\right)\e^{4A}
+b'\left(4\partial_\sigma^4 A_1 - 16 \partial_\sigma^2 A_1
\right) \nn
&& - 4(b+b')\left(\partial_\sigma^4 A_1 + 2 \partial_\sigma^2 A_1 
 - 6 (\partial_\sigma A_1)^2\partial_\sigma^2 A_1 \right), \\
\label{eq2pc}
0&=&-{l^4 \over 8\pi G}\e^{4A}\partial_z\phi
+ {4 \over 3}ab' \left(4\partial_\sigma^4 A_1 
- 16 \partial_\sigma^2 A_1\right) \ .
\eea
Then one gets  
\bea
\label{SP4}
0&=&{1 \over \pi G l}\left\{\sqrt{1 + {kl^2 \over 3R^2} 
+ {l^2 c^2 \over 24 R^8}} -1 \right\}R^4 + 8 b', \\
\label{SP5}
0&=& - {c \over 8\pi G} + 32 ab' \ .
\eea
Note that for minimal spinor coupling the second equation 
does not have a solution.
Eq.(\ref{SP5}) can be solved with respect to $c$:
\be
\label{SP5b}
c=32\times 8\pi G a b',
\ee
but the boundary value $\phi_0$ of $\phi$ becomes a free 
parameter.  

We should also note that in the classical case that $b'=0$, 
there is no solution for (\ref{SP4}) and (\ref{SP5}). From 
Eq.(\ref{SP5}), we find $c=0$ if $b'=0$. Then if we put $c=0$ 
and $b'=0$ in (\ref{SP4}), there is no solution.

When the dilaton vanishes ($c=0$) and the brane is 
the unit sphere ($k=3$), the equation (\ref{SP4}) reproduces the 
result of ref.\cite{HHR} for ${\cal N}=4$ $SU(N)$ super Yang-Mills 
theory in case of the large $N$ limit where 
$b'$ is replaced by $-{N^2 \over 4(4\pi )^2}$: 
\be
\label{slbr3}
{R^3 \over l^3}\sqrt{1 + {R^2 \over l^2}}={R^4 \over l^4}
+ {GN^2 \over 8\pi l^3}\ .
\ee 

Let us define a function $F(R, c)$ as 
\be
\label{FRc}
F(R,c)\equiv {1 \over \pi G l}\left(\sqrt{1 + {kl^2 \over 3R^2} 
+ {l^2 c^2 \over 24 R^8}} -1 \right)R^4 \ ,
\ee
which appears in the r.h.s. in (\ref{SP4}). 

For the $k>0$ case, 
$F(R,c)$ has a minimum at $R=R_0$, where $R_0$ is defined by
\be
\label{min}
0={8kl^2 \over 3R_0^2} + {k^2 l^4 \over R_0^4}
 - {2l^2 c^2 \over 3 R_0^8}\ .
\ee
When $k>0$, there is only one solution for $R_0$. 
Therefore $F(R,c)$ in the case of $k>0$ (sphere case) 
is a monotonically increasing function of $R$ when 
$R>R_0$ and a decreasing function when $R<R_0$. 
Since $F(R,c)$ is clearly a monotonically increasing 
function of $c$, we find for $k>0$ and $b'<0$ case 
that $R$ decreases when $c$ increases if $R>R_0$, that is, 
the non-trivial dilaton makes the radius smaller. 
We can also find that there is no solution for $R$ in 
(\ref{SP4}) for very large $|c|$. 

We can consider the $k<0$ case. When $c=0$, there is no solution for 
$R$ in (\ref{SP4}). We can find, however, there is a solution if 
$|c|$ is large enough:
\be
\label{G8}
{|c| \over \pi G \sqrt{24}} > -8b'\ .
\ee

Hence, for constant bulk potential there is the possibility of 
quantum creation of a 4d de Sitter or a 4d hyperbolic 
brane living in 5d AdS bulk space. This occurs even for not exactly 
conformal invariant quantum brane matter. 
This finishes our consideration of quantum induced dilatonic brane-worlds 
in Einstein frame.

\section{Quantum induced dilatonic brane-worlds in string frame.}

We now transform the brane-world action in the Einstein frame (see  
(\ref{Stotal})) into the Jordan frame. 
If we consider the scale transformation
\be
\label{sc1}
g_{\mu\nu}\rightarrow \e^\rho g_{\mu\nu}\ ,
\ee
with the choice
\be
\label{sc3}
\e^{\left({D\over 2} -1\right)\rho}=\alpha\phi\ ,\quad 
(\mbox{$\alpha$ is a constant})\ , 
\ee
we find that the actions (\ref{SEHi}), 
(\ref{GHi}) and (\ref{S1dil}) are transformed as
\bea
\label{SEHiJ}
\SEH&=&{1 \over 16\pi G}\int d^5 x \sqrt{\gfv}\left(\alpha\phi 
R_{(5)}  + {4\alpha \over 3\phi}\nabla_\mu\phi\nabla^\mu \phi 
-{\alpha \over 2}\phi \nabla_\mu\phi\nabla^\mu \phi \right.\nn
&& \left. + \left({12 \over l^2}+\Phi(\phi)\right)\left(\alpha
 \phi\right)^{5 \over 3}\right), \\
\label{GHiJ}
\SGH&=&{1 \over 8\pi G}\int d^4 x \alpha\phi \sqrt{\gfr}
\nabla_\mu n^\mu \ , \\
\label{S1dilJ}
S_1&=& -{1 \over 16\pi G}\int d^4x \left(\alpha\phi
\right)^{4 \over 3}\sqrt{\gfr}\left(
{6 \over l} + {l \over 4}\Phi(\phi)\right)\ .
\eea 

\subsection{Bulk solution in the string frame}

In the bulk, the variation over $\phi$ gives the following 
equation of motion:
\bea
\label{sc7}
0&=& \alpha R_{(5)} - {4\alpha \over 3\phi^2}\partial_\mu\phi
\partial^\mu\phi - {\alpha \over 2}\partial_\mu\phi
\partial^\mu\phi + {5 \over 3}\left({12 \over l^2} + \Phi(\phi)
\right)\alpha^{5 \over 3} \phi^{2 \over 3} \nn
&& + \Phi'(\phi)\left(\alpha\phi\right)^{5 \over 3} 
 - {8\alpha \over 3}\nabla_\mu\left({1 \over \phi}
\partial^\mu\phi\right) 
+ \alpha \nabla_\mu\left( \phi \partial^\mu\phi\right)\ .
\eea
On the other hand, the variation over the metric $g^{\mu\nu}$ 
gives
\bea
\label{sc8}
0&=& - {1 \over 2}\left(\alpha\phi R_{(5)} 
+ {4\alpha \over 3\phi}\partial_\mu\phi\partial^\mu\phi 
 - {\alpha \over 2}\phi\partial_\mu\phi\partial^\mu\phi 
+ \left({12 \over l^2} + \Phi(\phi)\right) 
\left(\alpha\phi\right)^{5 \over 3} \right)g_{(5)\mu\nu} \nn
&& + \alpha\phi R_{(5)\mu\nu} - \alpha\nabla_\mu\partial_\nu\phi 
+ \alpha g_{(5)\mu\nu}\Box \phi \nn
&& + {4\alpha \over 3\phi}\partial_\mu\phi\partial_\nu\phi 
 - {\alpha \over 2}\partial_\mu\phi\partial_\nu\phi\ .
\eea
Thus, one gets the bulk equations of motion in string frame.
Using (\ref{sc8}), we have
\bea
\label{sc9}
0&=&-{3 \over 2}\left(\alpha\phi R_{(5)} 
+ {4\alpha \over 3\phi}\partial_\mu\phi\partial^\mu\phi 
 - {\alpha \over 2}\phi\partial_\mu\phi\partial^\mu\phi 
 \right) \nn
&& - {5 \over 2}\left({12 \over l^2} + \Phi(\phi)\right)
\left(\alpha\phi\right)^{5 \over 3} + 4\alpha\Box\phi\ .
\eea
Substituting (\ref{sc9}) into (\ref{sc7}) and (\ref{sc8}), 
one  obtains
\bea
\label{sc10}
0&=&\alpha\nabla_\mu\left(\phi\partial^\mu\phi\right) 
+ \Phi'(\phi)\left(\alpha\phi\right)^{5 \over 3}
\\
\label{sc11}
0&=&-\alpha\nabla_\mu\partial_\nu\phi
 - {\alpha \over 3}g_{(5)\mu\nu}\Box\phi 
+ \alpha\phi R_{(5)\mu\nu} \nn
&& + {1 \over 3}\left({12 \over l^2} + \Phi(\phi)\right)
\left(\alpha\phi\right)^{5 \over 3}
g_{(5)\mu\nu}
+ {4\alpha \over 3\phi}\partial_\mu\phi\partial_\nu\phi
 -{\alpha \over 2}\phi\partial_\mu\phi\partial_\nu\phi\ .
\eea

First, let us  consider $\Phi(\phi)=0$ case. 
In the Einstein frame, the solution is given by (\ref{curv2}). 
The metric $g_{(5)\mu\nu}^{\rm J}$ in the Jordan frame  
is obtained with the help of (\ref{sc1}) and (\ref{sc3}), or more explicitly
\bea
\label{curv3}
g_{(5)\mu\nu}^{\rm J}dx^\mu dx^\nu&=&
\left(\alpha\phi\right)^{-{2 \over 3}}\left(
f(y)dy^2 + y\sum_{i,j=0}^{3}\hat g_{ij}(x^k)dx^i dx^j\right) \nn
f&=&{l^2 \over 4y^2 \left(1 + { c^2l^2 \over 24 y^4}
+ {kl^2 \over 3 y}\right)} \nn
\phi&=&c\int dy {l \over 2\sqrt{y^6 \left(1 
+ { c^2l^2 \over 24 y^4}
+ {kl^2 \over 3 y}\right)}}\ .
\eea
One can check directly that the metric (\ref{curv3}) satisfies 
Eqs.(\ref{sc10}) and (\ref{sc11}). 
Although the classical bulk solution in the Einstein frame is equivalent to
the one in 
 the Jordan frame, the physical interpretation of the spacetime is 
changed due to the factor of $\left(\alpha\phi
\right)^{-{2 \over 3}}$. 
Since the transformation is conformal, the causal structure of 
the spacetime is not changed, especially the situation that 
there is a curvature singularity at $y=0$ is not changed. 
When $y\rightarrow \infty$, however, the spacetime is not 
asymptotically AdS but the metric behaves as
\be
\label{sc12}
g_{(5)\mu\nu}^{\rm J}dx^\mu dx^\nu\sim 
\left(-{\alpha cl \over 4}\right)^{-{2 \over 3}}\left(
{l^2 \over 4 y^{2 \over 3}}dy^2 + y^{7 \over 3}
\sum_{i,j=0}^{3}\hat g_{ij}(x^k)dx^i dx^j\right) \ .
\ee
If one defines a coordinate $z$ by
\be
\label{sc13}
z\equiv \left(-{\alpha cl \over 4}\right)^{-{1 \over 3}}
{3l \over 4}y^{2 \over 3}\ ,
\ee
the metric in (\ref{sc12}) is rewritten by
\be
\label{sc14}
g_{(5)\mu\nu}^{\rm J}dx^\mu dx^\nu\sim 
dz^2 + \left(-{\alpha cl \over 4}\right)^{{1 \over 2}}
\left(4z \over 3l\right)^{7 \over 2}
\sum_{i,j=0}^{3}\hat g_{ij}(x^k)dx^i dx^j \ .
\ee
Then the warp factor behaves as the power of $z$, instead of 
the exponential function in Einstein frame.

One can also consider the case that the dilaton potential 
${12 \over l^2} + \Phi(\phi)$ is given by (\ref{ts5}).  
Using the relation (\ref{sc1}) and (\ref{sc3}) between the 
Einstein frame and the Jordan frame, from (\ref{ts4}) and 
(\ref{ts6}), we find the following 
solution:
\bea
\label{sc15}
g_{(5)\mu\nu}^{\rm J}dx^\mu dx^\nu&=&
\left(\pm\alpha\sqrt{6}\ln (m^2 y)\right)^{-{2 \over 3}}
\left({1 \over -{2ky \over 9} + {f_0 \over y^2}}
dy^2 + y\sum_{i,j=0}^{3}\hat g_{ij}(x^k)dx^i dx^j\right) \nn
\phi&=&\pm \sqrt{6} \ln (m^2 y)\ .
\eea
One can again check that the above solution satisfies 
Eqs.(\ref{sc10}) and (\ref{sc11}). 
Then the above result is equivalent with that in the Einstein 
frame. Comparing the obtained metric with that in the Einstein 
frame in (\ref{ts4}) and (\ref{ts6}), there appears the 
factor of the logarithmic function of $y$, coming from the 
conformal transformation. In other words, the interpretation of lenghts 
in both frames is different while solutions are equivalent. 

\subsection{Brane solutions in the string frame}

Having proof of explicit equivalency of bulk solutions, one can 
analyze the brane. From the actions in 
(\ref{SEHiJ}), (\ref{GHiJ}) and (\ref{S1dilJ}), the variation 
over $\phi$ gives the following equation on the boundary
\bea
\label{Jb1}
0&=&{l^4\e^{4A} \over 8\pi G}\left\{\left({8\alpha \over 3\phi_0}
 - \alpha\phi_0\right)\partial_z\phi + 8\alpha\partial_z A 
\right.\nn
&& \left. - {4\alpha \over 3}
 \left({6 \over l} + {l \over 4}\Phi(\phi)\right)
\left(\alpha\phi_0\right)^{1 \over 3}
 - {l \over 4}\Phi'(\phi)\left(\alpha\phi\right)^{4 \over 3}
 \right\}\ .
\eea
Here we choose the metric as in (\ref{metric1}) and $\phi_0$ is 
the value of $\phi$ on the boundary.  
The variation over $A$ gives the following equation 
\be
\label{Jb2}
0={48 l^4 \over 16\pi G}\e^{4A}\left(\alpha\phi_0\partial_z A 
+ {\alpha \over 3}\partial_z\phi - {1 \over 6}
 \left({6 \over l} + {l \over 4}\Phi(\phi)\right)
\left(\alpha\phi_0\right)^{4 \over 3}\right)\ .
\ee
The coordinate $z$ and $A$ in the warp factor are related with 
those in the Einstein frame, $z_E$ and $A_E$ by 
\be
\label{Jb3}
dz_E=\left(\alpha\phi\right)^{1 \over 3}dz\ ,\quad
A_E=A+{1 \over 3}\ln \left(\alpha\phi\right)\ .
\ee
Then Eqs.(\ref{Jb1}) and (\ref{Jb2}) are rewritten as
\bea
\label{Jb4}
0&=&{l^4\e^{4A_E} \over 8\pi G}\left\{-\partial_{z_E}\phi 
+ \alpha\left(\alpha\phi_0\right)\left(
8\partial_{z_E} A_E
 - {4\alpha \over 3}
 \left({6 \over l} + {l \over 4}\Phi(\phi)\right)
\right.\right.\nn
&& \left.\left. - {l \over 4}\Phi'(\phi)\right) \right\}\\
\label{Jb5}
0&=& {48 l^4 \over 16\pi G}\e^{4A_E}\left\{\partial_{z_E}A_E 
 - {1 \over 6} \left({6 \over l} 
 + {l \over 4}\Phi(\phi)\right)\right\}\ .
\eea
Combining (\ref{Jb4}) and (\ref{Jb5}), we obtain
\be
\label{Jb6}
0={l^4\e^{4A_E} \over 8\pi G}\left\{-\partial_{z_E}\phi 
 - {l \over 4}\Phi'(\phi) \right\}\ .
\ee
The obtained equations (\ref{Jb5}) and (\ref{Jb6}) are identical 
with the corresponding equations (\ref{eq2b}) and (\ref{eq2pb}) 
without the quantum correction, respectively. 

Choosing the metric of 5 dimensional space-time as in 
(\ref{metric1}):
\be
\label{metric1b}
ds^2=dz^2 + \e^{2A(z,\sigma)}\tilde g_{\mu\nu}dx^\mu dx^\nu\ ,
\quad \tilde g_{\mu\nu}dx^\mu dx^\nu\equiv l^2\left(d \sigma^2 
+ d\Omega^2_3\right)\ ,
\ee
where $d\Omega^2_3$ corresponds to the metric of 3 dimensional 
unit sphere, we now include the quantum correction as 
in (\ref{W2}): 
\bea
\label{W2b}
W&=& b \int d^4x \sqrt{\widetilde g}\widetilde F A \nn
&& + b' \int d^4x\left\{A \left[2{\widetilde\Box}^2 
+\widetilde R_{\mu\nu}\widetilde\nabla_\mu\widetilde\nabla_\nu 
 - {4 \over 3}\widetilde R \widetilde\Box^2 
+ {2 \over 3}(\widetilde\nabla^\mu \widetilde R)\widetilde\nabla_\mu
\right]A \right. \nn
&& \left. 
+ \left(\widetilde G - {2 \over 3}\widetilde\Box \widetilde R
\right)A \right\} \\
&& -{1 \over 12}\left\{b''+ {2 \over 3}(b + b')\right\}
\int d^4x \left[ \widetilde R - 6\widetilde\Box A 
 - 6 (\widetilde\nabla_\mu A)(\widetilde \nabla^\mu A)
\right]^2 \ .\nonumber
\eea
Note that as typically in Jordan frame there is no non-minimal dilaton
coupling with matter we took minimal spinors, i.e.
 $a=0$. 
Then one obtains the following brane equations 
(instead of (\ref{Jb1}) and (\ref{Jb2})):
\bea
\label{Jb7}
0&=&{l^4\e^{4A} \over 8\pi G}\left\{\left({8\alpha \over 3\phi_0}
 - \alpha\phi_0\right)\partial_z\phi + 8\alpha\partial_z A 
\right.\nn
&& \left. - {4\alpha \over 3}
 \left({6 \over l} + {l \over 4}\Phi(\phi)\right)
\left(\alpha\phi_0\right)^{1 \over 3}
 - {l \over 4}\Phi'(\phi)\left(\alpha\phi\right)^{4 \over 3}
 \right\} \nn
&& +b'\left(4\partial_\sigma^4 A - 16 \partial_\sigma^2 A
\right) \nn
&& - 4(b+b')\left(\partial_\sigma^4 A + 2 \partial_\sigma^2 A 
 - 6 (\partial_\sigma A)^2\partial_\sigma^2 A \right), \\
\label{Jb8}
0&=&{48 l^4 \over 16\pi G}\e^{4A}\left(\alpha\phi_0\partial_z A 
+ {\alpha \over 3}\partial_z\phi - {1 \over 6}
 \left({6 \over l} + {l \over 4}\Phi(\phi)\right)
\left(\alpha\phi_0\right)^{4 \over 3}\right) \nn
&& + {4 \over 3}ab' \left(4\partial_\sigma^4 A 
- 16 \partial_\sigma^2 A\right) \ .
\eea
For $\Phi(\phi)=0$ case,  substituting the solution 
in (\ref{curv3}), one finds
\bea
\label{Jb9} 
0&=&{1 \over \pi G l}\left\{\sqrt{1 + {kl^2 \over 
3\left(\alpha\phi_0\right)^{2 \over 3}R^2} 
+ {l^2 c^2 \over 24 \left(\alpha\phi_0\right)^{8 \over 3}R^8}}
 -1 \right\}\left(\alpha\phi_0\right)^{4 \over 3}R^4 \nn
&& + 8 b', \\
\label{Jb10}
0&=& - {c \over 8\pi G}  \nn
&& + {1 \over \pi G l\phi_0}\left\{\sqrt{1 + {kl^2 \over 
3\left(\alpha\phi_0\right)^{2 \over 3}R^2} 
+ {l^2 c^2 \over 24 \left(\alpha\phi_0\right)^{8 \over 3}R^8}}
 -1 \right\}\left(\alpha\phi_0\right)^{4 \over 3}R^4\ .
\eea
Combining (\ref{Jb9}) and (\ref{Jb10}), one gets
\be
\label{Jb11}
0= - {c \over 8\pi G}  - {8b' \over \phi_0}\ .
\ee
Eq.(\ref{Jb11}) has non-trivial solution and can 
be solved with respect to $\phi_0$:
\be
\label{Jb12}
\phi_0=-{64\pi G b' \over c}\ .
\ee
In the classical case that $b'=0$, 
there is no solution for (\ref{Jb9}). Let us define a 
function $F(R, c)$ as 
\be
\label{FRcb}
F(R,c)\equiv 
{1 \over \pi G l}\left\{\sqrt{1 + {kl^2 \over 
3\left(\alpha\phi_0\right)^{2 \over 3}R^2} 
+ {l^2 c^2 \over 24 \left(\alpha\phi_0\right)^{8 \over 3}R^8}}
 -1 \right\}\left(\alpha\phi_0\right)^{4 \over 3}R^4 \ ,
\ee
It appears in the r.h.s. in (\ref{Jb9}). 

For  $k>0$ case, 
$F(R,c)$ has a minimum at $R=R_0$, where $R_0$ is defined by
\be
\label{minb}
0={8kl^2 \over 3\left(\alpha\phi_0\right)^{2 \over 3}R_0^2} 
+ {k^2 l^4 \over \left(\alpha\phi_0\right)^{4 \over 3}R_0^4}
 - {2l^2 c^2 \over 3 \left(\alpha\phi_0\right)^{8 \over 3}R_0^8}\ .
\ee
When $k>0$, there is only one solution for $R_0$. 
Therefore $F(R,c)$ in the case of $k>0$ (sphere case) 
is a monotonically increasing function of $R$ when 
$R>R_0$ and a decreasing function when $R<R_0$. 
Since $F(R,c)$ is clearly a monotonically increasing 
function of $c$, we find for $k>0$ and $b'<0$ case 
that $R$ decreases when $c$ increases if $R>R_0$, that is, 
the non-trivial dilaton makes the radius smaller. 
 
Since one finds  
\be
\label{F1}
F(R_0,c)={kl \left(\alpha\phi_0\right)^{2 \over 3}R_0^2 
\over 4\pi G},
\ee
using (\ref{FRcb}) and (\ref{minb}), 
Eq.(\ref{Jb9}) has a solution if 
\be
\label{F2}
{kl \left(\alpha\phi_0\right)^{2 \over 3}R_0^2 
\over 4\pi G}\leq -8b'\ .
\ee
That puts again some bounds to the dilaton value.
When $|c|$ is small,  using (\ref{minb}), one obtains 
\be
\label{F3}
R_0^4\sim {2c^2\left(\alpha\phi_0\right)^{-{4 \over 3}} 
\over 3k^2 l^2}\ ,\quad 
F(R_0,c)\sim {1 \over 4\pi G}{|c| \over \sqrt{3}}\ .
\ee
Therefore Eq.(\ref{F2}) is satisfied for small $|c|$. 
On the other hand, when $c$ is large, we get 
\be
\label{F4}
R_0^6\sim {c^2\left(\alpha\phi_0\right)^{-{6 \over 3}} 
\over 4k}\ ,\quad 
F(R_0,c)\sim {\left(k |c| \right)^{2 \over 3} \over 
4^{4 \over 3}\pi G}\ .
\ee
Therefore Eq.(\ref{F2}) is not always satisfied and 
we have no solution for $R$ in (\ref{SP4}) for very 
large $|c|$. 

We now consider the $k<0$ case. When $c=0$, there is no 
solution for $R$ in (\ref{Jb9}). Let us define another 
function $G(R,c)$ 
as follows:
\be
\label{G1}
G(R,c)\equiv 1 + {l^2 c^2 \over 24 
\left(\alpha\phi_0\right)^{8 \over 3}R^8} 
+ {kl^2 \over 3\left(\alpha\phi_0\right)^{2 \over 3}R^2}\ .
\ee
Since $G(R,c)$ appears in the root of $F(R,c)$ in (\ref{FRcb}), 
$G(R,c)$ must be positive. Then since 
\be
\label{G2}
{\partial G(R,c) \over \partial R}=-{l^2 c^2 \over 
3\left(\alpha\phi_0\right)^{8 \over 3}R^9}
- {2kl^2 \over 3\left(\alpha\phi_0\right)^{2 \over 3}R^3}\ ,
\ee
$G(R,c)$ has a minimum 
\be
\label{G3}
1+{kl^2 \over 4}\left(-{2k \over c^2}\right)^{1 \over 3},
\ee
when 
\be
\label{G4}
R^6 = -{c^2\left(\alpha\phi_0\right)^{-{6 \over 3}} \over 2k}\ .
\ee
Therefore if 
\be
\label{G5}
c^2\geq {k^4 l^6 \over 32}\ ,
\ee
$F(R,c)$ is real for any positive value of $R$. Since 
\be
\label{G6}
F(0,c)={|c| \over \pi G \sqrt{24}},
\ee
and when $R\rightarrow \infty$
\be
\label{G7}
F(R,c)\rightarrow {kl \left(\alpha\phi_0\right)^{2 \over 3}R^2 
\over 6\pi G}<0\ ,
\ee
there is a solution $R$ in (\ref{Jb9}) if 
\be
\label{G8b}
{|c| \over \pi G \sqrt{24}} > -8b'\ .
\ee
This is the same bound as in Einstein frame (previous section).

Thus we demonstrated the complete equivalency of quantum induced 
inflationary (hyperbolic) dilatonic brane-worlds in Einstein and string
(Jordan) frames.
 
Note that 
Eq.(\ref{Jb9}) is identical with the corresponding equation 
(\ref{SP4}) in the Einstein frame if we regard 
$\left(\alpha\phi_0\right)^{1 \over 3}R$ as the radius $R_E$ 
in the Einstein frame:
\be
\label{Jb13}
R=\left(\alpha\phi_0\right)^{-{1 \over 3}}R_E\ .
\ee
Then the solution has properties similar to those 
in the Einstein frame. 
Since $b'$ is order $N$ quantity from (\ref{SP3}), 
Eq.(\ref{Jb12}) and (\ref{Jb13}) might tell that the radius $R$ 
in the Jordan frame is much smaller than the radius $R_E$ in the 
Einstein frame if $N$ is large. In case that the brane is sphere, 
the brane becomes de Sitter space.  Since the rate 
of the expansion is given by ${1 \over R}$ in  de Sitter 
space, the rate might become much larger if compare with that in the 
Einstein frame when $N$ is large. Thus, even having formal equivalency, 
the physical interpretation of results obtained in Jordan and Einstein 
frames may be  different.

\section{Brane-world black holes in string and Einstein frames}

%%%%%%%%%%%

In analogy with Randall-Sundrum model \cite{RS},
we now consider the following classical action of the gravity
coupled with dilaton $\phi$ in the Einstein frame with 
Lorentzian signature:
\bea
\label{S}
S&=& {1 \over 16\pi G}\left[ \int d^5 x
 \sqrt{-g_{(5)}}\left( R_{(5)}
 - {1 \over 2}\partial_\mu\phi \partial^\mu\phi
 - V(\phi) \right) \right. \nn
&& \left. - \sum_{i={\rm hid},{\rm vis}}
\int_{B_i} d^4 x \sqrt{-g_{(4)}} U_i(\phi)\right]\ .
\eea
Here $B_{\rm hid}$ and $B_{\rm vis}$ are branes corresponding
to hidden and visible sectors respectively and 
$U_i(\phi)$ corresponds to the vacuum energies
on the branes in \cite{RS}.   One assumes $U(\phi)$ is
dilaton dependent and its form is explicitly given later on from
the consistency of the equations of motion.
The dilaton potential $V(\phi)$ is often given in terms of 
the superpotential $W(\phi)$ :
\be
\label{Vi}
V=\left({\partial W \over \partial \phi}\right)^2
 - {4 \over 6 } W^2 \ .
\ee

We assume again $\phi$ only depends on $z$ and 
the metric has the following form:
\be
\label{Mi2}
ds^2=dz^2 + \e^{2A(z)}\tilde g_{ij}dx^i dx^j\ .
\ee
Here $\tilde g_{ij}$ is the metric of the Einstein manifold. 
We also suppose the hidden and
visible branes sit on $z=z_{\rm hid}$ and $z=z_{\rm vis}$,
respectively. Then the equations of motion are given by
\bea
\label{Ei}
&& \phi''+ 4A'\phi' = {\partial V \over \partial \phi}
+ \sum_{i={\rm hid},{\rm vis}}
{\partial U_i(\phi) \over \partial \phi} \delta(z-z_i)\ , \\
\label{Eii}
&& 4A''+ 4(A')^2 + {1 \over 2}(\phi')^2 \nn
&& \quad = - {1 \over 3}V(\phi) - {2 \over 3}\sum_{i={\rm hid},{\rm vis}}
U_i(\phi)\delta(z-z_i) \ , \\
\label{Eiiib}
&& A'' + 4 (A')^2 = k\e^{-2A} - {1 \over 3}V(\phi)
 - {1 \over 6}\sum_{i={\rm hid},{\rm vis}}
U_i(\phi)\delta(z-z_i) \ .
\eea
Here $'\equiv {d \over dz}$.
Especially when $k=0$, Eqs. (\ref{Ei}-\ref{Eiiib}) have the
following first integrals in the bulk:
\bea
\label{Iii}
\phi'=\sqrt{2}{\partial W \over \partial \phi}\ ,
\quad A' = - {1 \over 3\sqrt{2}}W\ .
\eea
Near the branes,
Eqs. (\ref{Ei}-\ref{Eiiib}) have the following form :
\be
\label{Eiv}
\phi'' \sim {\partial U_i(\phi)\over \partial\phi}\delta (z-z_i)\ ,
\quad A'' \sim -{U_i(\phi) \over 6}\delta (z-z_i)\ ,
\ee
or
\be
\label{Eivb}
2\phi' \sim {\partial U_i(\phi)\over \partial\phi} ,
\quad 2A' \sim -{U_i(\phi) \over 6}\ ,
\ee
at $z=z_i$.
Comparing (\ref{Eivb}) with (\ref{Iii}), we find
\be
\label{Ev}
U_{\rm hid}(\phi)= 2\sqrt{2}W(\phi)\ ,\quad
U_{\rm vis}(\phi)=- 2\sqrt{2}W(\phi)\ .
\ee
We should note that $k=0$ does not always mean the brane is flat.
As well-known, the Einstein equations are given by,
\be
\label{A1}
R_{\mu\nu}-{1 \over 2}g_{\mu\nu}R+{1 \over 2}\Lambda g_{\mu\nu}
= T^{\rm matter}_{\mu\nu}\ .
\ee
Here $T^{\rm matter}_{\mu\nu}$ is the energy-momentum tensor of
the matter fields. If we consider the vacuum solution where
$T^{\rm matter}_{\mu\nu}=0$, Eq.(\ref{A1}) can be rewritten as 
\be
\label{A2}
R_{\mu\nu}={\Lambda \over 2}g_{\mu\nu}\ .
\ee
If we put $\Lambda=2k$, Eq.(\ref{A2}) is nothing but the equation
for the Einstein manifold. The Einstein manifolds
are not always homogeneous manifolds like flat Minkowski,
(anti-)de Sitter space
\be
\label{AdS}
ds_4^2= -V(r)dt^2 + V^{-1}(r)dr^2+r^2 d\Omega^2,
\qquad
V(r) = 1 - \frac{\Lambda}{6} r^2,
\ee
or Nariai space
\be
\label{Nsol}
ds_4^2={1 \over \Lambda}\left( \sin^2 \chi d\psi^2
- d\chi^2 - d\Omega^2\right)\ .
\ee
but they can be some
black hole solutions like Schwarzschild-(anti-)de Sitter
black hole
\be
\label{SAdS}
ds_4^2= -V(r )dt^2 + V^{-1}(r )dr^2+r^2 d\Omega^2,
\qquad
V(r) = 1- {\tilde G_4 M \over r} - \frac{\Lambda}{6} r^2\ .
\ee
As a special case, one can also consider $k=0$ solution like
Schwarzschild black hole,
\be
\label{schw}
ds_4^2\equiv \tilde g_{ij}dx^i dx^j
=-\left(1 - {\tilde G_4 M \over r}\right)dt^2
+{dr^2 \over \left(1 - {\tilde G_4 M \over r}\right)}
+ r^2 d\Omega^2\ .
\ee
In (\ref{SAdS}) and (\ref{schw}),
$M$ is the mass of the
black hole on the brane and the effective gravitational
constant $G_4$ on the 3-brane (here $d=4$)
is given by
\be
\label{schw1}
{1 \over G_4}
={1 \over G}\int_{z_{\rm hid}}^{z_{\rm vis}} dz \e^{(d-2)A}\ .
\ee
In these solutions, the curvature singularity
at $r=0$ has a form of line penetrating the bulk 5d universe and
the horizon makes a tube surrounding the singularity.
The singularity and the horizon connect the hidden and visible
branes. These black holes have been discussed in ref.\cite{NOOO}.

We now consider  the Jordan frame, in order to see if
singularity supports (or breaks) the equivalency on classical level.
Using scale transformation given by (\ref{sc1}) and (\ref{sc3}) 
with $D=5$, the action (\ref{S}) is rewritten as
\bea
\label{S2}
S&=&{1 \over 16\pi G}\int d^5 x \sqrt{\gfv}\left(\alpha\phi 
R_{(5)}  + {4\alpha \over 3\phi}\nabla_\mu\phi\nabla^\mu \phi 
-{\alpha \over 2}\phi \nabla_\mu\phi\nabla^\mu \phi \right.\nn
&& \left. - V(\phi)\left(\alpha \phi\right)^{5 \over 3}\right) \nn
&& \left. - \sum_{i={\rm hid},{\rm vis}}
\int_{B_i} d^4 x \sqrt{-g_{(4)}} \left(\alpha\phi
\right)^{4 \over 3}U_i(\phi)\right]\ .
\eea
Then if we choose the metric as in (\ref{Mi2}) in the Jordan 
frame and  $\phi$ only depends on $z$ again, we obtain the 
following equations instead of (\ref{Ei}), (\ref{Eii}) and 
(\ref{Eiiib}), 
\bea
\label{EiJ}
&& \alpha\left(\phi\phi''+ 4A'\phi\phi' 
+ \left(\phi'\right)^2\right) \nn
&& = {\partial V \over \partial \phi}
\left(\alpha\phi\right)^{5 \over 3}
+ \sum_{i={\rm hid},{\rm vis}}
{\partial U_i(\phi) \over \partial \phi} 
\left(\alpha\phi\right)^{4 \over 3} \delta(z-z_i)\ , \\
\label{EiiJ}
&& \alpha\phi\left(4A''+ 4(A')^2\right) 
+ {\alpha \over 2}\phi(\phi')^2 - {4\alpha \over 3\phi}(\phi')^2 
+ {4\alpha \over 3}\left(\phi''+ A'\phi'\right)
\nn
&& \quad = - {1 \over 3}V(\phi)\left(\alpha\phi\right)^{5 \over 3}
 - {2 \over 3}\sum_{i={\rm hid},{\rm vis}}
U_i(\phi)\left(\alpha\phi\right)^{4 \over 3}\delta(z-z_i) \ , \\
\label{EiiibJ}
&& \alpha\phi\left(A'' + 4 (A')^2\right) + {\alpha \over 3}\left(\phi''
+ 7A'\phi'\right) \nn
&& = k\alpha\phi\e^{-2A}
 - {1 \over 3}V(\phi)\left(\alpha\phi\right)^{5 \over 3}
 - {1 \over 6}\sum_{i={\rm hid},{\rm vis}}
U_i(\phi)\left(\alpha\phi\right)^{4 \over 3}
\delta(z-z_i) \ .
\eea
If one transforms the above equations to those in the Einstein 
frame by changing 
\bea
\label{Tr}
A&\rightarrow& A - {1 \over 3}\ln \left(\alpha\phi\right) \nn
dz&\rightarrow& \left(\alpha\phi\right)^{-{1 \over 3}}dz \nn
&& \left(\begin{array}{rcl}
'\equiv \partial_z &\rightarrow & \left(\alpha\phi
\right)^{1 \over 3}\partial_z \\
''=\partial_z^2 &\rightarrow & \left(\alpha\phi
\right)^{2 \over 3}\left(\partial_z^2 + {\partial_z\phi 
\over3\phi}\partial_z\right) 
\end{array}
\right)\ ,
\eea
then Eqs.(\ref{Ei}), (\ref{Eii}) and (\ref{Eiiib}), which are the 
corresponding equations in the Einstein frame, are reproduced. 
Thus we can confirm the equivalence between the Jordan frame 
and the Einstein frame  description of dilatonic brane-world black holes 
on the classical level. Their physical interpretation may 
be again different.

\section{Discussion}

In summary, we discussed AdS/CFT induced quantum dilatonic 
brane-worlds where branes may be flat, de Sitter (inflationary) 
or Anti-de Sitter Universe. Actually, such objects appear 
in frames of AdS/CFT correspondence \cite{AdS} as warped 
compactification of relevant holographic RG flow \cite{NOZ,HHR}.
The role of free parameter (brane tension) is played by 
effective brane tension produced by conformal anomaly of QFT 
sitting on the brane. Hence, only brane quantum effects are 
considered. We compared the construction of such quantum 
dilatonic brane-worlds in two frames: string and Einstein one. 
The very nice feature of brane-worlds is discovered: in all 
examples under consideration the string and Einstein
frames are eqiuvalent! This holds to be true also for the number 
of classical dilatonic brane-world black holes. This is 
completely different from the case of quantum corrected 4d 
dilatonic gravity (section 2) where de Sitter Universe with 
decaying dilaton exists in Einstein frame but does not exist 
in Jordan frame. 

Quantum effects may be useful in other aspects of brane-worlds. 
In particulary, for flat branes the bulk quantum effects 
(Casimir force) may be estimated \cite{GPT,NOZ1,HKP} and used 
for radion stabilization. Unfortunately, in usual Randall-Sundrum 
Universe such quantum effects are actually supporting the radion 
destabilization. Nevertheless, in the case of thermal
Randall-Sundrum scenario \cite{BMNO} such quantum effects may 
not only stabilize the radion but also may provide the necessary 
mass hierarchy \cite{BMNO} (at least, for some temperatures). 
It would be extremely interesting to estimate the bulk quantum 
effects for dilatonic backgrounds and to understand 
their role (as well as frame dependence of such Casimir effect) in
the creation of dilatonic brane-worlds.

Another interesting line of research is related with account 
of quantum effects on graviton perturbations around the brane. 
As is demonstrated in previous section, they may modify the 
massive graviton modes around hyperbolic brane. Clearly, in 
other regimes for quantum induced dilatonic (asymptotically) 
AdS brane more complicated dynamics may be expected.

\section*{Acknowledgements}

We thank  J. Socorro for participation at the early stage of 
this work. SDO is grateful to L. Randall for useful discussion. 
The work by O.O., S.D.O. and V.I.T. has been supported 
in part by CONACyT grant 28454E and that by S.D.O. in part 
by CONACyT(CP, ref.990356).

\appendix

\section{Remarks on gravitational perturbations around 
hyperbolic brane}

In \cite{KR,KMP}, the AdS$_4$ branes in AdS$_5$ were discussed 
and the existence of the massive normalizable mode of graviton 
was found. In these papers, the tensions of the branes are free 
parameters but in the case treated in the present paper, the 
tension is dynamically determined. 

Let us study the role of dynamically generated tension in getting of
massive graviton modes. Moreover, we consider dilatonic brane-world.
We now regard the brane as an object with a tension $U(\phi)$ 
and assume the brane can be effectively described by the 
folowing action:
\be
\label{bten1}
S_{\rm brane}=- {1 \over 16\pi G}\int d^4x \sqrt{-g_{(4)}}
U(\phi)\ .
\ee
If one assumes the metric in the form of (\ref{metric1}), then 
 using the Einstein equation, we find
\be
\label{bten2}
\partial_z^2 A + 4\left(\partial_z A\right)^2 
=k\e^{-2A} + {4 \over l^2} + {\Phi(\phi) \over 3} 
 - {U(\phi) \over 6}\delta(z-z_0) \ .
\ee
Then at $z=z_0$, 
\be
\label{bten3}
\left. \partial_z A \right|_{z=z_0}=-{U(\phi) \over 12}\ .
\ee
For simplicity, we consider the case of the constant dilaton 
potential $\Phi(\phi)=0$. Comparing (\ref{bten3}) 
with (\ref{eq2c}) and (\ref{SP4}), one gets 
\be
\label{bten4}
U(\phi) = - {12 \over l} + {96\pi G b' \over R^4}\ .
\ee
We should note that the tension becomes $R$ dependent due to 
the quantum correction. In case of AdS brane $k<0$, if no 
 dilaton is included, 
the boundary equation (\ref{SP4}) does not have any solution 
for $R$. When there is non-trivial dilaton and the parameter 
$c$ is large enough, Eq.(\ref{SP4}) has a solution. If 
$c$ is very large
\be
\label{lR}
R^4 \sim {c \over \pi G} + 8b'\ .
\ee

We now consider the perturbation by assuming the metric 
in the following form:
\be
\label{bten5}
ds^2=\e^{2\hat A(\zeta}\left(d\zeta^2 + \left(\hat g_{\mu\nu}
+ \e^{-{3 \over 2}\hat A(\zeta)}h_{\mu\nu}\right) dx^\mu dx^\nu\right)\ .
\ee
By choosing the gauge conditions 
$h^\mu_{\ \mu}=0$ and $\nabla^\mu h_{\mu\nu}=0$, 
one obtains the following equation
\be
\label{bten7}
\left(-\partial_\zeta^2 + {9\over 4} \left(\partial_\zeta \hat A
\right)^2 + {3 \over 2}\partial_\zeta^2 \hat A \right)h_{\mu\nu}
= m^2 h_{\mu\nu}
\ee
Here $m^2$ corresponds to the mass of the graviton on the brane 
\be
\label{bten8}
\left(\hat\Box \pm {1 \over R^2}\right)h_{\mu\nu} = m^2 
h_{\mu\nu}\ .
\ee
Here $\hat\Box$ is 4- dimensional d'Alembertian constructed on  
$\hat g_{\mu\nu}$ and the $+$ ($-$) sign corresponds to 
(anti-)de Sitter brane. 
Since $-\e^{A}d\zeta = dz = \sqrt{f}dy$ and 
$\e^A={\sqrt{y} \over l}$, 
we find, especially for the case of the constant dilaton 
potential, 
\be
\label{bten10}
\zeta=-\int dy \sqrt{f(y) \over y} 
= - {l^2 \over 2} \int {dy \over \sqrt{y^3\left(1 
+ {c^2 l^2 \over 24 y^4} + {kl^2 \over 3y}\right)}} \ .
\ee
We now consider the case that $c$ is very large, then
\be
\label{bten11}
f(y)\sim {6y^2 \over c^2}\ .
\ee
Since $y_0=R^2$ if there is a brane at $y=y_0$, Eq.(\ref{lR}) 
can be rewritten as 
\be
\label{lR2}
y_0^2 \sim {c \over \pi G} + 8b'\ .
\ee
If we choose $\zeta=0$ when $y=y_0$, Eqs.(\ref{bten10}) and 
(\ref{bten11}) give
\be
\label{bten12}
|\zeta|=-{1 \over |c|}\sqrt{8 \over 3}y^{3 \over2} + \zeta_0\ ,
\quad \zeta_0\equiv {1 \over |c|}\sqrt{8 \over 3}y_0^{3 \over2}
>0\ .
\ee
Note that the brane separates two bulk regions corresponding to 
$\zeta<0$ and $\zeta>0$, respectively.
Since $y$ takes the value in $[0,y_0]$, $\zeta$ takes the 
value in $[-\zeta_0,\zeta_0]$. 
Since $A={1 \over 2}\ln y$, from (\ref{bten7}), one gets
\be
\label{bten13}
\left(-\partial_\zeta^2 - {1 \over 4\left(|\zeta|
-\zeta_0\right)^2} - {1 \over \zeta_0}\delta(\zeta)
\right)h_{\mu\nu}
= m^2 h_{\mu\nu}
\ee
The zero mode solution with $m^2$ of (\ref{bten13}) is given by
\be
\label{bten14}
h_{\mu\nu}=\sqrt{\zeta_0 -|\zeta|}\ .
\ee
The general solution of (\ref{bten13} with $m^2\neq 0$ is given 
by the Bessel functions:
\be
\label{bten15}
h_{\mu\nu}=aJ_0\left(m\left(\zeta_0 - |\zeta|\right)\right)
+ b N_0\left(m\left(\zeta_0 - |\zeta|\right)\right)\ .
\ee
The coefficients $a$ and $b$ are constants of the integration and they 
are determined to satisfy the boundary condition 
\be
\label{bten16}
\left.{\partial_\zeta h_{\mu\nu} \over h_{\mu\nu}}
\right|_{\zeta\rightarrow 0+}
=-{1 \over 2\zeta_0}\ .
\ee
Note that  zero mode solution (\ref{bten14}) satisfies 
this boundary condition (\ref{bten16}). If $b\neq 0$, the 
solution in (\ref{bten15}) diverges at $\zeta=\pm\zeta_0$ and 
would not be normalizable. 
If $b=0$, the condition (\ref{bten16}) reduces to
\be
\label{bten17}
J_1(m\zeta_0)=0\ ,
\ee
that is 
\be
\label{bten18}
m\zeta_0=0,\,3.8317...,\,7.0155...,\, \cdots\ .
\ee
The non-vanishing solutions for $m^2$ give the mass of the 
massive graviton modes. Thus, these results indicate that 4d dilatonic
gravity 
on quantum induced hyperbolic brane may be trapped near the brane. 

Since $\zeta_0$ is given by $y_0$ in (\ref{bten12}) and $y_0$ is 
expressed by (\ref{lR2}), with the help of $b'$, which comes from the quantum 
correction and is negative, the quantum correction makes $\zeta_0$ 
smaller and increases the massive graviton mode mass $m$.
It would be of interest to discuss graviton/dilaton perturbations around 
asymptotically hyperbolic brane in other regimes and to compare the
corresponding predictions in different frames.

\end{document}